\documentclass[aps,floats,twocolumn,showpacs,preprintnumbers,nofootinbib]{revtex4}
\usepackage{pslatex}
\usepackage[T1]{fontenc}
\usepackage[latin1]{inputenc}
\usepackage{graphicx}
\usepackage{epsfig}
\usepackage{amsmath}

\usepackage{ifthen}

\newcommand{\lsim}{\mbox{\raisebox{-.6ex}{~$\stackrel{<}{\sim}$~}}}
\newcommand{\gsim}{\mbox{\raisebox{-.6ex}{~$\stackrel{>}{\sim}$~}}} 

\newcommand{\be}{\begin{equation}}
\newcommand{\ee}{\end{equation}}
\newcommand{\bea}{\begin{eqnarray}}
\newcommand{\eea}{\end{eqnarray}}
\newcommand{\mg}{m_{3/2}}
\newcommand{\K}{K\"{a}hler\hspace{0.1cm}}
\newcommand{\OR}{O'Raifeartaigh\;}
\newcommand{\g}{{3/2}}
\newcommand{\Hub}{H}
\newcommand{\mpl}{M^2_{Pl}}
\newcommand{\half}{\frac{1}{2}}

\newcommand{\nc}{\newcommand}
\nc{\renc}{\renewcommand}
\nc{\eqs}[2]{\mbox{Eqs.~(\ref{#1},\,\ref{#2})}}
\nc{\eq}[1]{\mbox{Eq.~(\ref{#1})}}
\nc{\figs}[2]{\mbox{Figs.~(\ref{#1},\,\ref{#2})}}
\nc{\fig}[1]{\mbox{Fig~.(\ref{#1})}}

\nc{\bee}[1]{\begin{equation} \mbox{$\label{#1}$}}
\nc{\eee}{\vspace{0.1cm}\end{equation}}
 \def\gae{\; ^{>}_{\sim} \;}

\def\eV{\mathrm{\ eV}}
\def\GeV{\mathrm{\ GeV}}
\def\MeV{\mathrm{\ MeV}}
\def\keV{\mathrm{\ keV}}
\def\TeV{\mathrm{\ TeV}}

\def\lsim{\;^{<}_{\sim} \;} \def\gsim{\; ^{>}_{\sim} \;}

\begin{document}
\preprint{HIP-2009-24/TH}

\title{Cosmological evolution of scalar fields and gravitino dark matter in gauge mediation at low reheating temperatures}
\author{Andrea Ferrantelli}
\email{andrea.ferrantelli@helsinki.fi}
\affiliation{University of Helsinki and Helsinki Institute of Physics, P.O.Box 64, FIN-00014 University of Helsinki, Finland}
\author{John McDonald}
\email{j.mcdonald@lancaster.ac.uk}
\affiliation{Cosmology and Astroparticle Physics Group, University of 
Lancaster, Lancaster LA1 4YB, UK}

\begin{abstract}

        We consider the dynamics of the supersymmetry-breaking scalar field and the production of dark matter gravitinos via its decay in a gauge-mediated supersymmetry breaking model with metastable vacuum. We find that the scalar field amplitude and gravitino density are extremely sensitive to the parameters of the hidden sector. For the case of an O'Raifeartaigh sector, we show that the observed dark matter density can be explained by gravitinos even for low reheating temperatures $T_{R} \lsim 10 \GeV$. Such low reheating temperatures may be implied by detection of the NLSP at the LHC if its thermal freeze-out density is in conflict with BBN.

  \end{abstract}

\maketitle

\section{Introduction} 

        With the advent of the LHC, it is possible that in the not-too-distant future we will confirm supersymmetry (SUSY) as the next level of particle physics. Moreover, based on the pattern of superpartner masses, we may be able to conclude that SUSY breaking is of gauge-mediated type \cite{GMSB, GR}. In this case it becomes probable that dark matter is due to gravitino LSPs \cite{gravitino}. Gravitinos can be generated in a variety of ways; via thermal scatterings \cite{ellis, BBB, PS, Andrea, Narendra}, via inflaton \cite{infldecay} and moduli decay \cite{moduli,Dine}, and in the case of GMSB models, via decay of the energy density in the SUSY breaking hidden sector \cite{Coughlan, Banks, IK, ET, HKT}. In all cases a fundamental parameter is the reheating temperature $T_{R}$. Thermal gravitino dark matter generally requires a large $T_{R} > 10^{6} \GeV$ \cite{Kaz}. Production via inflaton decay, on the other hand, is strongly dependent on the branching ratio of inflaton decay to gravitinos \cite{Felder, GTA, Maroto, Kawasaki}. In the case of production via decay of a scalar field in the GMSB sector, the gravitino density will depend on both the initial amplitude of the scalar field oscillations and on $T_{R}$.

      It is generally not possible to obtain direct information about $T_{R}$ from cosmological observations. However, it may perhaps be possible to deduce $T_{R}$ from information obtained at the LHC. The 
point here is that Big-Bang Nucleosynthesis (BBN) can place strong constraints on the NLSP \cite{Kaz,Jedamzik}. With enough information about NLSP couplings, it is possible to calculate the thermal freeze-out density of NLSPs and to compare this with BBN constraints. If the thermal freeze-out density is too large, the simplest explanation in the context of the MSSM is that the thermal freeze-out NLSP density is diluted by a low reheating temperature below its freeze-out temperature. In this case we would need a mechanism to generate the gravitino density with a low reheating temperature, $T_{R} \lsim 10 \GeV$. In the following we will consider whether such a low reheating temperature can be consistent with gravitinos from the GMSB sector.  

     The SUSY breaking sector generally requires a gauge singlet field, in order to give masses to gauginos. At the end of inflation, its scalar component starts coherent oscillations around the minimum of its potential, due to the expansion of the Universe. A huge amount of energy is accordingly released, and typically a large spectrum of particles is produced. This mechanism created much interest immediately after its discovery \cite{T}, as it can lead to cosmological difficulties which are related to either BBN \cite{BBN} or to the overproduction of gravitinos \cite{gravover}. The latter comes not only from the thermal production through scatterings in the primordial bath, but in gauge mediation also from the decays of particles into gravitinos, which in this class of model are the LSPs. 
			
       In this paper we investigate the possibility of dark matter gravitinos from the metastable SUSY breaking hidden sector, in particular for the case of a low reheating temperature \cite{Lowreh}. In order to be able to explicitly calculate the gravitino density for all field amplitudes without cut-off, we focus on the case of a perturbative \OR hidden sector. In particular, we focus on models of gauge mediation with metastable vacua, which have been discussed by Murayama and Nomura in \cite{MN}. 

			In this framework, gravitino production has been recently discussed by several authors (for instance, in \cite{IK}, \cite{ET} and \cite{HKT}). It was found that the right amount of dark matter gravitinos can be produced by the decay of the supersymmetry breaking field. Here we perform a study in the same direction, though there are two important differences. First, in \cite{ET} the minimum $S_c$ of the scalar potential when the potential is dominated by Hubble corrections is independent of the Hubble scale and therefore constant in time. Once the $S$ mass term dominates the potential, the field begins coherent oscillations with initial amplitude $S_c$. On the contrary, we show in Section \ref{dynamics} that the dynamics is quite different. The minimum of the potential when Hubble corrections are important is time dependent and evolves towards the time-independent minimum. The $S$ field tracks this minimum until Hubble corrections are small and then enters coherent oscillations. This dramatically alters the resulting gravitino density. Our main results are (i) the coherent oscillation amplitude and gravitino density are extremely sensitive to the parameters of the hidden sector and (ii) the reheating temperature can easily be as low as the BBN lower bound, $T_{R} \approx 1 \MeV$, and still be consistent with gravitino dark matter from the decay of the GMSB scalar. 

			The inclusion of an \OR hidden sector is the second feature of this work, since it is the first time that gravitino production in GMSB with metastable vacua has been studied using a completely perturbative model rather than an effective model with a cutoff. In the latter case, the gravitino abundance and $T_R$ are dependent on this arbitrary cutoff (\cite{IK, ET, HKT}). By introducing an explicit hidden sector of the \OR type, both $\Lambda$ and the reheating temperature are explicitly determined as functions of the parameters of the theory. This means that the model at hand contains only quantities, such as the coupling constants, which can be directly constrained via cosmological observations or at colliders, since they determine the mass spectrum in the observable sector.

			The paper is organised as follows. In Section \ref{dynamics} we study the dynamics of the SUSY breaking field $S$ and the amplitude of its coherent oscillations. In Section \ref{cosmo} we calculate the gravitino number density
and the required reheating temperature under the assumption that gravitinos are the principal constituents of dark matter.
In Section IV we present our conclusions.  In the Appendix we calculate quantum and supergravity corrections to the scalar potential $V(S)$ and derive bounds on the parameters of the model and on the gravitino mass range.

\section{Coherent oscillations of the SUSY breaking scalar field in an expanding background}\label{dynamics}

\begin{figure}
\includegraphics[scale=1.4]{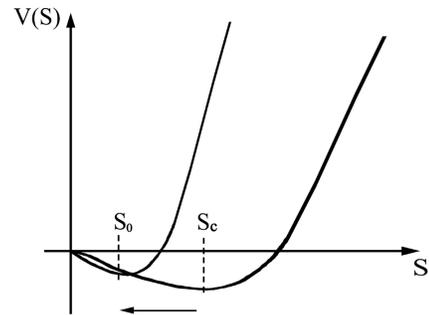}
\caption{Time-dependent shift of the minimum of the potential due to the expansion of the Universe.}
\label{fig:shift}
\end{figure}

\begin{figure}
\includegraphics[scale=1.4]{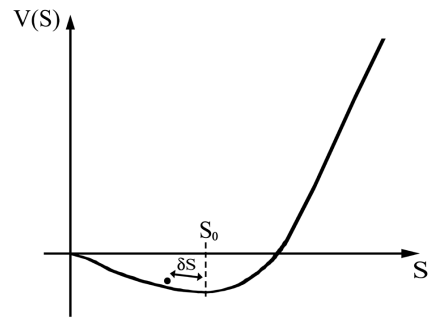}
\caption{The displacement from the minimum due to the time evolution of $S_0$.}
\label{fig:deltaS}
\end{figure}

As already pointed out in the Introduction, our study of the dynamics of the $S$-field differs from what is assumed in the literature \cite{ET}, where the initial amplitude $S_c$ of the scalar field oscillations is assumed to be determined by the Hubble corrections (with the \K potential normalized in units of the Planck mass),
\be
\Delta V \approx e^K(3H^2)\approx 3H^2\left[|S|^2-\frac{5kM}{16\pi^2}(S+S^{\dagger})+\dots  \right],
\label{h2corr}
\ee
which imply $|S_c|=5kM/16\pi^2$. This assumes that the inflaton dominates the energy density of the Universe when the $S$ field starts oscillating. We will also include this contribution in our analysis. However, in addition to (\ref{h2corr}), the scalar potential contains terms which are not $H$-dependent. As we will see, this implies that the minimum $S_0(t)$ is time-dependent in the presence of Hubble corrections. The $S$ field tracks the minimum as it decreases towards the time-independent ($H = 0$) minimum. This generates a displacement of $S$ from the minimum, as shown in Fig.\ref{fig:shift}.
As $H$ decreases, either the quadratic or the linear term becomes dominated by the $H=0$ factor. So the minimum of $V$ is $H$-dependent until \emph{both} $|S|^2$ and $(S+S^\dagger)$ become $H$-independent. The displacement
 from the minimum at this time, $\delta S$, then determines the initial amplitude of the $S$ oscillations about the time-independent minimum $S_0$, as shown in Fig.\ref{fig:deltaS}.

	This scenario assumes that the $S$ field is not thermalized\footnote{The case of a thermalized hidden sector was discussed in \cite{therm}.}. For $T_{R}$ less than the messenger mass $M$, the scattering of $S$ particles by thermal MSSM particles will be due to messenger loops, with rate $\Gamma_{s} \approx \alpha_{s} T^5/M^4$, where $\alpha_{s}$ accounts for coupling and numerical factors. Comparing with the expansion rate at $T < T_{R}$, $H \approx T^2/M_{Pl}$, the hidden sector will not be thermalized as long as $T_{R} \lsim (M/\alpha_{s}M_{Pl})^{1/3}M$. We will see later that this is easily satisfied in the models of interest to us here.

  We first discuss the dynamics of the GMSB sector generally, then we will apply it to the specific case of the model defined in Section \ref{cosmo}. Let us consider a potential of the general form
\be
V=\frac{1}{2}m_S^2S^2-aS+3H^2\left(\frac{S^2}{2}-cS\right),
\label{genpot}
\ee
where the scalar field is assumed to take real values without loss of generality (with $|S| \rightarrow S/\sqrt{2}$ for canonical normalization). When $H$ decreases from a large value, either the linear or the quadratic term becomes dominated by the $H$-independent term. We define $H_1$ and $H_2$ by 
\bea
&&H=H_1 \qquad \mathrm{when} \qquad a=3cH^2,\\
&&H=H_2 \qquad \mathrm{when} \qquad m_S^2=3H^2.
\eea

\begin{figure}
\includegraphics[scale=1.4]{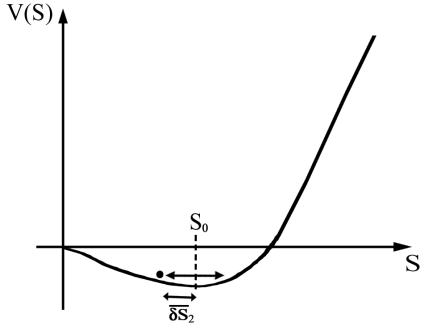}
\caption{Evolution of $\delta S$ in the potential with minimum at $\overline{\delta S_2}$.}
\label{fig:deltaS2}
\end{figure}

In the case $H_2>H_1$, the quadratic term becomes $H$-independent first. The potential when $H_2 > H > H_1$ is then 
\be
V\approx\frac{1}{2}m_S^2S^2-3H^2cS. 
\ee
The minimum is then $S_0(t)=3cH^2/m_S^2$. The field $S$ will track this time-dependent minimum. We define $S=S_0+\delta S$, where $\delta S$ is a perturbation. Then the scalar field equation for $S$, 
\be
\ddot{S}+3H\dot{S}=-\frac{\partial V}{\partial S},
\ee
becomes\footnote{We have checked that the higher-order terms in the expansion of the potential are negligible throughout our analysis.}
\be
\delta \ddot{S}+3H\delta \dot{S}=-(\ddot{S}_0+3H\dot{S}_0)-\left.\frac{\partial V}{\partial S}\right|_{S_0}
-\left.\frac{\partial^2 V}{\partial S^2}\right|_{S_0}\delta S.
\label{fried}
\ee
If the background is dominated by the inflaton, the Hubble parameter $H$ depends on the scale factor $a(t)$ as $H\propto a^{-3/2}$. Thus the above equation takes the form
\be
\delta\ddot{S}+3H\delta\dot{S}=-\frac{27}{2}\frac{c H^4}{m_S^2}-m_S^2\delta S,
\ee
and the displacement $\delta S$ evolves as if it had a potential with minimum given by
\be
\overline{\delta S_2}=-\frac{27}{2}\frac{c H^4}{m^4_S}.
\label{S2}
\ee
This is illustrated in Fig.\ref{fig:deltaS2}.

On the other hand, if $H_1>H_2$ then it is the linear term which becomes $H$-independent first. The potential when 
$H_1 > H > H_2$ is then 
\be
V\approx \frac{3}{2}H^2S^2-aS.
\ee
The minimum is at $S_0=a/3H^2$. In this case Eq.(\ref{fried}) becomes
\be
\delta\ddot{S}+ \delta\dot{S}=-\frac{9a}{2}- 3 H^2 \delta S,
\ee
which implies 
\be
\overline{\delta S_1}=-\frac{3a}{2H^2}.
\label{S1}
\ee
The shifts (\ref{S2}) and (\ref{S1}) give the initial amplitude of the oscillations of $S$ about the $H$-independent minimum
in the two cases, where oscillations begin once $H \approx H_1$ and $H \approx H_2$ respectively.

\section{Decay of the SUSY breaking field into gravitinos and constraints on $T_R$}\label{cosmo}

\subsection{A model of GMSB with metastable vacua and a hidden sector of the \OR type}

    We consider a superpotential \cite{MN},
\be
W=-\mu^2S+\lambda SX^2+mXY+kSf\bar{f}+Mf\bar{f},
\label{super}
\ee
and a non-minimal \K potential for the SUSY breaking field $S$,
\be
K=|S|^2-\frac{|S|^4}{4\Lambda^2}+\mathcal{O}\left( \frac{|S|^6}{\Lambda^4} \right).
\label{kp}
\ee
The above equations define a model of gauge mediation with metastable vacua. In particular, there is a negative quartic term for $S$ in the \K potential generated by radiative corrections, an accidental $U(1)_R$ symmetry which is broken in the messenger sector, and an explicit mass term for the messengers. As shown in Appendix A, the breaking of the additional symmetry induces a linear term in the \K potential, which makes the SUSY breaking field deviate from the origin. Such an effect, coupled to the expansion of the Universe, is responsible for the oscillations of the field $S$ and accordingly, for its decay into gravitinos.

    The superpotential written in Eq.(\ref{super}) corresponds to an \OR hidden sector which contains, besides $S$, the singlet fields $X$ and $Y$. These have an approximately canonical \K potential (up to terms suppressed by the Planck mass). The parameters  $\lambda$ and $k$ can be taken real and positive without loss of generality, and the scales $\mu$, $m$, and $M$ have dimension of mass. The messenger sector contains Standard Model multiplets belonging to ${\bf 5} + {\bf 5^*}$ representations of $\rm SU(5)$, and the messengers $f$ and $\bar{f}$ are assumed to have a canonical \K potential as well. Due to the second term in (\ref{kp}), a positive mass $m_S=\mu^2/\Lambda$ for $S$ is generated around the origin. The gravitino mass range allowed by the metastable vacua is in principle very large, $1\eV\lsim\mg\lsim 10\GeV$ \cite{MN}. However, it is shown in Appendix B that it is possible to constrain this interval, by virtue of the explicit \OR hidden sector and phenomenological constraints.

As discussed in Appendix A and shown in \cite{MN}, radiative corrections generate $\Lambda$ as a function of the mass scale $m$ and coupling $\lambda$ in the hidden sector,
\be
\Lambda^2=\frac{3\pi^2m^2}{\lambda^4}\,.
\label{cutoff}
\ee
Along the direction $X=0$, $Y=0$, Eqs.(\ref{super}) and (\ref{kp}) localize the global SUSY minimum for the scalar potential at
\be
S=-\frac{M}{k}\,, \hspace{1cm} f=\bar{f}=\frac{\mu}{\sqrt{k}}\,.
\ee
Moreover, to avoid tachyonic messengers, we require that the condition
\be
M^2>k\mu^2
\ee
is satisfied. In this case, we find a local minimum at $S=f=\bar{f}=0$, which is a metastable vacuum. At this point in field space supersymmetry is broken, with $F_{S} = \mu^2$. As we will see shortly, $S$ is a flat direction at the classical level, but it is lifted around the origin by the radiative corrections to the \K potential that are generated by the loops of the fermions and scalars in the hidden sector and of the messengers. The mass matrix in the basis of the $X$ and $Y$ fermions $\psi_X$ and $\psi_Y$ is
\be
  M^h_\psi=\left( \begin{array}{cc}
    2\lambda S & m \\
    m          & 0
  \end{array} \right),
\label{fmm}
\ee
while the $F$-terms of the chiral superfields, in the basis $(X,X^{\dagger},Y,Y^{\dagger})$, produce the following scalar mass matrix \cite{MN}
\be
M^h_\phi=\left( \begin{array}{cccc}
    m^2 + 4\lambda^2 |S|^2  & -2\lambda\mu^2  &  2\lambda m S^\dagger  &  0 \\
    -2\lambda\mu^2  & m^2 + 4\lambda^2 |S|^2  &  0  &  2\lambda m S         \\
    2\lambda m S     &     0   &  m^2  &  0 \\
    0  &  2\lambda m S^\dagger &  0  &  m^2
  \end{array} \right) ~.
\label{smm}
\ee
For the messenger sector, it can be shown that the scalar and fermion mass matrices $M^m_\phi$ and $M^m_\psi$, in the bases $(f,f^{\dagger},\bar{f},\bar{f}^{\dagger})$ and $(\psi_f,\psi_{\bar{f}})$, can be respectively written as
\bea
\lefteqn{
 \hspace{-2mm}M^m_\phi = } \nonumber \\ &&\hspace{-1cm} \left(\hspace{-2mm}
 \begin{array}{cccc}
    |M+kS|^2  & 0  &  0  &  -k\mu^2+k\lambda X^{\dagger 2} \\
    0  & |M+kS|^2 &  -k\mu^2+k\lambda X^2  &  0 \\
    0  & -k\mu^2+k\lambda X^{\dagger 2}   &  |M+kS|^2  &  0 \\
     -k\mu^2+k\lambda X^2 & 0 &  0  &  |M+kS|^2
  \end{array} \hspace{-1mm}\right)\nonumber\\
\label{msmm}
\eea
and
\be
  M^m_\psi=\left( \begin{array}{cc}
    0 &M+kS \\
    M+kS & 0
  \end{array} \right).
\label{fmmm}
\ee
The radiative corrections generated by these particles shift the minimum of the scalar potential only slightly from the origin. However, the shift due to these terms is important from a cosmological point of view, since it is responsible for the production of gravitinos through the decay of the $S$ field, as it is shown in the following.

\subsection{Gravitino dark matter and reheating temperature}

      The results of Section \ref{dynamics} and the GMSB model summarised above can be now applied to gravitino production. Here we calculate the gravitino number density as a function of the reheating temperature $T_R$ and the masses and couplings of the \OR sector.
The scalar potential of the field $S$ is obtained from the superpotential (\ref{super}) and from the \K potential (\ref{kp}), as a sum of a classical part and the corrections generated by quantum loops and by supergravity. Each of these contributions is discussed in detail in the Appendix.
In particular, the one-loop corrections \cite{CW} to the \K potential can be calculated by using the following expression \cite{Grisaru}:
\bea
&&K^{(1)}=-\frac{1}{32\pi^2}\mathrm{Tr}\left[\bar{\mu}\mu\ln\frac{\bar{\mu}\mu}{\Lambda^2}-2M\ln\frac{M}{\Lambda^2}\right]=\nonumber\\
&=&-\frac{1}{32\pi^2}\left\{\mathrm{Tr}\left[M^2_\phi\left(\ln\frac{M^2_\phi}{\Lambda^2}-1\right)\right]-\right.\nonumber\\
&&\left.-2\mathrm{Tr}\left[M^2_V\left(\ln\frac{M^2_V}{\Lambda^2}-1\right)\right]\right\}
\label{Kcorr}
~,\eea
which is generally valid for a renormalizable $N=1$ theory. Here $M^2_\phi$ and $M^2_V$ are the mass matrices of the chiral and vector superfield sectors respectively, with
\be
(M^2_\phi)_{ij}=\overline{W}_{i\bar{k}}\delta^{\bar{k}k}W_{kj} \qquad \textrm{or} \qquad \mu_{ik}=m_{ik}+\lambda_{ijk}\Phi^j,
\label{masscorr}
\ee
for the chiral sector and
\be
M_{AB}=\half\left(\overline{X}^i_A X_{Bi}+\overline{X}^i_B X_{Ai}\right),
\ee
for the vector sector (which is however absent in the model discussed here).
By summing equations (\ref{potglob}) and (\ref{sugrafinal}), we obtain
\bea
&V(S)= V_{SUSY}+V_{Sugra}\approx\nonumber\\
&\approx \left(3H^2+\dfrac{2\lambda^4\mu^4}{3\pi^2m^2}+\dfrac{5k^6\mu^8}{48\pi^2M^6} \right)|S|^2+\nonumber\\
&+\dfrac{5 k}{16 \pi^2} M(S+S^\dagger)\left(-3H^2-\dfrac{2\mu^4}{\mpl}+\dfrac{2k^2\mu^4}{M^2}\right)
\label{total}
~,\eea
where we have kept only the linear and quadratic terms in $S$, in the direction corresponding to $X=Y=f=\bar{f}=0$. These terms are dominant for sufficiently small values of the field.

	The scalar field can be treated as a modulus, with interactions given in \cite{IK}. The decay into gravitinos is not helicity suppressed, and gravitino production from $S$ decay dominates production from inflaton decay\footnote{The mass scales here considered are consistent with the discussion in Ref.\cite{Dine}.}.

Dominant decay into a pair of gravitinos ($S\rightarrow\psi_\mu\psi_\nu$) occurs if the cutoff $\Lambda$ satisfies \cite{ET}
\be
\Lambda \lsim
8 \times 10^{14} \sqrt{\frac{\alpha_3}{0.1}}\left(\frac{m_3}{1\TeV} \right)^{-1}\left(\frac{\mg}{1\GeV}\right)\;\GeV
\label{Lambdaconstr}
~.\ee
The inclusion of the \OR sector makes it possible to relate this directly to the mass scale $m$ in the hidden sector
using Eq.(\ref{cutoff}), 
\be m \lsim 1.5 \times 10^{14} \; \lambda^2 \GeV\times\sqrt{\frac{\alpha_3}{0.1}}\left(\frac{m_3}{1\TeV} \right)^{-1}\left(\frac{\mg}{1\GeV} \right)\,
\label{mdecay}
~.\ee
This corresponds to an upper bound in the range $10^8\GeV - 10^{15}\GeV$ for a gravitino mass between $1\keV$ and $10\GeV$ (the lower bound on $m_{3/2}$ is discussed in Appendix B).

     The messenger mass scale $M$ is related to the gluino mass $m_3$ and to the SUSY breaking scale $\mu$ by \cite{ET}
\bee{m3}  M = \frac{\alpha_{3} k \mu^{2}}{4 \pi m_{3}}    ~.\eee 
Therefore, by using the relation between $\mu$ and $m_{3/2}$, 
\bee{mu2} \mu^2 = \sqrt{3} m_{3/2} M_{Pl}   ~,\eee
with the reduced Planck mass $M_{Pl}=2.43\times 10^{18}\GeV$, we find
\be
M \approx 3.3 \times 10^{13} \; k \left(\frac{\alpha_3}{0.1} \right)\left(\frac{1\TeV}{m_3} \right)\left(\frac{\mg}{1\GeV} \right) \; \GeV    \,.
\label{massM}
\ee
This implies $10^{7}\GeV\lsim M \lsim 10^{14}\GeV$ for the above range of gravitino mass. For this range of messenger mass, the corresponding upper bound on the reheating temperature for which the $S$ field is not thermalized is $1 \TeV$ to $10^{12} \GeV$ (using $\alpha_{s}^{1/3} \sim 1$).

	Eq.(\ref{massM}), together with (\ref{mdecay}), gives the following approximation for the coefficient of the linear term in the scalar potential (\ref{total}),
\be
\left( -3 H^2 -\dfrac{2 \mu^{4}}{\mpl}+\dfrac{2k^2 \mu^{4}}{M^2}\right) \approx \left(-3H^2 + 
\dfrac{2k^2 \mu^{4}}{M^2}\right),
\label{num}
\ee
over the entire gravitino mass range considered. Similarly, for the coefficient of the quadratic term, 
\be
\left(3H^2+ \dfrac{2\lambda^4\mu^4}{3\pi^2m^2}+\dfrac{5k^6\mu^8}{48\pi^2M^6}\right)\approx\left(3H^2+ \dfrac{2\lambda^4\mu^4}{3\pi^2m^2}\right).
\label{quadr}
\ee

    One should also consider what happens when adding a constant term $\sim \mg\mpl$ to the superpotential of the model. This is needed to tune the vacuum energy to zero in supergravity \footnote{The authors acknowledge Kazunori Kohri for this idea.}. Recalling Eq.(\ref{super}), we can write \cite{IK}
\be
W=-\mu^2S+\mg\mpl+...  ~\;\;\;\;.
\ee
This generates both linear and quadratic terms in $S$ through the factor $|W|^2$ in the supergravity corrections \cite{Nilles},
\bea
&V_{Sugra}  =  \exp(K/\mpl)  \left[ \sum_{\alpha,\beta}
\left( \dfrac{\partial^2 K}{\partial \bar{\phi}_{\alpha} \partial \phi_{\beta} }
	\right)^{-1}\times\right.\nonumber\\
	 &\left.\times
\left( \dfrac{\partial W}{\partial \phi_{\alpha} } + \dfrac{W}{\mpl}
	 \dfrac{\partial K}{\partial \phi_{\alpha} } \right)
	 \left( \dfrac{\partial \overline{W} }{\partial \bar{\phi}_{\beta} }
	+ \dfrac{\overline{W}}{\mpl}
	\dfrac{\partial K}{\partial \bar{\phi}_{\beta} } \right)
- 3 \dfrac{|W|^2}{\mpl}  \right]\,.\nonumber\\
\eea
In fact, the contribution of $\mg\mpl$ is \cite{Lyth},
\bea
&V_{Sugra}= \exp(K/\mpl) \left(...-3 \dfrac{|W|^2}{M_{Pl}^2} \right)\approx\nonumber\\
&\approx\left(1+\dfrac{K}{\mpl} \right)6\mu^2\mg(S+S^{\dagger})\approx\nonumber\\
&\approx\left[1+\dfrac{|S|^2}{\mpl}-\dfrac{5k}{16\pi^2}\dfrac{M}{\mpl}(S+S^{\dagger})\right]6\mu^2\mg(S+S^{\dagger})\nonumber\\
&\approx2\sqrt{3}\dfrac{\mu^4}{M_{Pl}}(S+S^{\dagger})-\dfrac{15k}{4\sqrt{3}\pi^2}\dfrac{M\mu^4}{M_{Pl}^3}|S|^2\,.
\eea
	Eqs.(\ref{num}) and (\ref{quadr}) change accordingly:
\bea
&\left( 3H^2 -\dfrac{2k^2\mu^4}{M^2}-\dfrac{32\sqrt{3}\pi^2\mu^4}{5k}\dfrac{1}{M M_{Pl}}
\right)\,,
\label{lin_corr}\\
&\left(3 H^{2} + \dfrac{2\lambda^4\mu^4}{3\pi^2m^2}-\dfrac{15k\mu^4}{4\sqrt{3}\pi^2}\dfrac{M}{M_{Pl}^3}\right)\,.
\label{quad_corr}
\eea
For the entire gravitino mass range here considered, the contribution of the additional term in (\ref{quad_corr}) is always negligible. On the other hand, if $k\lsim 0.1$ the new term in (\ref{lin_corr}) dominates for certain values of $\mg$, thus we take it into account as well.

	The scalar potential of interest is therefore the following,
\bea
&V(S)\approx \left(3H^2+\dfrac{2\lambda^4\mu^4}{3\pi^2m^2}\right)\dfrac{S^2}{2}+\nonumber\\
&+\dfrac{5 kM}{8 \pi^2} \dfrac{S}{\sqrt{2}}\left(-3H^2+\dfrac{2k^2\mu^4}{M^2}+\dfrac{32\sqrt{3}\pi^2\mu^4}{5k}\dfrac{1}{M M_{Pl}}\right)
\label{potential}
,\eea
where it is assumed that $S$ is real for simplicity, which will be the case on minimizing the potential if all parameters are considered real. Equation (\ref{potential}) can thus be rewritten as follows,
\bea
&V(S)\approx \dfrac{\lambda^4\mu^4}{3\pi^2m^2}S^2+\left(\dfrac{5 k^3\mu^4}{4 \sqrt{2} \pi^2M}+2\sqrt{6}\dfrac{\mu^4}{M_{Pl}} \right)S+\nonumber\\
&+3H^2\left(\dfrac{S^2}{2}-\dfrac{5 kM }{8 \sqrt{2} \pi^2}S \right).
\label{recast}
\eea
Comparison of this expression with the potential
\be
V=\frac{1}{2}m_S^2S^2-aS+3H^2\left(\frac{S^2}{2}-cS\right)\,,
\ee
gives the parameters which determine the dynamics of the field $S$, as discussed in Section \ref{dynamics}.

     We next review the calculation of the gravitino abundance from decay of a scalar field $S$. Oscillations of $S$ start when the Universe is still dominated by the inflaton, thus $T>T_R$. Let $n_{S\;osc}$ be the number density of the scalar and $a_{osc}$ the scale factor at that time. The number density $n_S(T)$ at a generic temperature is given by
\bea
n_S(T)&=&\left( \frac{a_{osc}}{a(T)} \right)^3 n_{S\;osc}  = \left( \frac{a_{osc}}{a(T_R)} \right)^3\left( \frac{a(T_R)}{a(T)} \right)^3 n_{S \; osc} =\nonumber\\
&=&\left( \frac{\Hub(T_R)}{H_{osc}} \right)^2 \frac{g(T)T^3}{g(T_R)T_R^3} n_{S\;osc}\,,
\eea
since $H\propto a^{-3/2}$ during inflaton domination. Recalling the expression for the Hubble scale during radiation domination, the above becomes
\be
n_S(T)=\frac{\pi^2 }{90}\frac{g(T) T^3T_R}{\mpl}\frac{n_{S\;osc}}{H_{osc}^2}\,.
\ee
Therefore the number density to entropy density of $S$ is 
\be
\frac{n_S}{s}=\frac{T_R}{4\mpl}\frac{n_{S\;osc}}{H_{osc}^2}\,.
\ee
From the scalar potential Eq.(\ref{recast}), $\rho_S = m_S^2S^2 /2$, which implies $n_{S \;osc}=\rho_S/m_S=(1/2)m_SS_{osc}^2$, where $S_{osc}$ is the value of the field at the beginning of the oscillations. Therefore
\be
\frac{n_S}{s} = \frac{m_S T_R S_{osc}^2}{8\mpl H_{osc}^2}\,.
\label{yieldosc}
\ee
Since each $S$ scalar decays into a pair of gravitinos, the gravitino number density to entropy ratio is given by  
\be
\frac{n_{3/2}}{s} = \frac{m_S T_R S_{osc}^2}{4\mpl H_{osc}^2}\,
\label{yieldgrav}
~,\ee
where the initial oscillation amplitude around the minimum, $S_{osc}$, is given by $\overline{\delta S_1}$ or $\overline{\delta S_2}$ at the time when all $H$ dependence becomes negligible.

	Consider first the case where $H_2>H_1$, namely where $H_{osc}^2 \approx H_{1}^2 = -a/3c$ and $S_{osc} \equiv \overline{\delta S_2}$. 
Therefore
\be
\frac{n_{S}}{s} = \frac{m_S T_R}{8\mpl}\frac{\overline{\delta S_2}^2}{H_1^2}\,.
\label{yielddelta}
\ee
The parameters $a$ and $c$ and the mass $m_S$ are fixed by the messenger and \OR hidden sectors,
\bea
&&m_S^2=\dfrac{2\lambda^4\mu^4}{3\pi^2m^2}\,,\\
&&a= -\dfrac{5 k^3\mu^4}{4 \sqrt{2} \pi^2M}\left(1+\dfrac{16\sqrt{3}\pi^2}{5k^3}\dfrac{M}{M_{Pl}}\right)\,,\\
&&c = \frac{5  k M}{8 \sqrt{2}\pi^2}\,.
\eea
Using these we obtain 
\be
\dfrac{n_{S}}{s} \approx \frac{2025 \pi^7}{16 \sqrt{2}} \frac{k^4 m_{3}^4 m^{7}T_R}{\lambda^{14} \alpha_{3}^4 m_{3/2}^5 M_{Pl}^7}\left[1+\dfrac{12\pi\alpha_3}{5k^2}\left(\dfrac{\mg}{m_3} \right) \right]^3 \,,
\label{yieldS}
\ee
where we have used \eq{m3} and \eq{mu2} to eliminate $M$ and $\mu$. 
Therefore  
\bea
&& \dfrac{n_{3/2}}{s} \approx 0.11 \times \; \dfrac{k^4}{\lambda^{14}}\left(\dfrac{\alpha_3}{0.1}\right)^{-4}\left(\dfrac{m_3}{1\TeV}\right)^{4}\left(\dfrac{\mg}{1\GeV} \right)^{-5} \nonumber\\
&&\times\left(\dfrac{m}{10^{14}\GeV}\right)^7\left(\dfrac{T_R}{10^8\GeV}\right)\left[1+\dfrac{12\pi\alpha_3}{5k^2}\left(\dfrac{\mg}{m_3} \right) \right]^3\,,\nonumber\\
\eea
and 
\bea
&&\Omega_{3/2}h^2 \approx 
1.7\times 10^{7} \; \dfrac{k^4}{\lambda^{14}}\left(\dfrac{\alpha_3}{0.1}\right)^{-4}\left(\dfrac{m_3}{1\TeV}\right)^{4}\left(\dfrac{\mg}{1\GeV} \right)^{-4} \nonumber\\
&&\times\left(\dfrac{m}{10^{14}\GeV}\right)^7\left(\dfrac{T_R}{10^8\GeV}\right)\left[1+\dfrac{12\pi\alpha_3}{5k^2}\left(\dfrac{\mg}{m_3} \right) \right]^3\,.\nonumber\\
\label{gnd}
\eea
By demanding that the gravitino is the principal constituent of dark matter,
\be
\Omega_{3/2}h^2\approx\mathcal{O}(0.1)\,,
\label{cdm}
\ee
we then obtain the reheating temperature $T_R$ as a function of the parameters of the model
\bea
&T_R\approx 0.6 \GeV \times \dfrac{\lambda^{14}}{k^4}\left(\dfrac{\alpha_3}{0.1}\right)^{4}
\left(\dfrac{m_3}{1\TeV}\right)^{-4}
\nonumber\\
&\times\left(\dfrac{\mg}{1\GeV} \right)^{4}\left(\dfrac{m}{10^{14}\GeV}\right)^{-7}\left[1+\dfrac{12\pi\alpha_3}{5k^2}\left(\dfrac{\mg}{m_3} \right) \right]^{-3}
\label{reheating21}
\eea

     In the case $H_1>H_2$, $H_{osc}^2 \approx H_{2}^{2} = m_{S}^{2}/3$ and $S_{osc} \equiv \overline{\delta S_1}$.   
We then obtain
\be
\frac{n_{S}}{s} \approx \frac{18225 \pi^3}{256 \sqrt{2}} \frac{k^4 m_{3}^2  m^{5} T_R}{\lambda^{10} \alpha_{3}^2  m_{3/2}^3 M_{Pl}^5}\left[1+\dfrac{12\pi\alpha_3}{5k^2}\left(\dfrac{\mg}{m_3} \right) \right]^2 \,.
\label{yieldS1}
\ee
Therefore 
\bea
& \dfrac{n_{3/2}}{s}  \approx 3.9\times10^{-3}\dfrac{k^4}{\lambda^{10}}\left(\dfrac{\alpha_3}{0.1}\right)^{-2}\left(\dfrac{m_3}{1\TeV}\right)^2\left(\dfrac{\mg}{1\GeV}\right)^{-3}
\nonumber\\
&\times\left(\dfrac{m}{10^{14}\GeV}\right)^5\left(\dfrac{T_R}{10^8\GeV}\right)\left[1+\dfrac{12\pi\alpha_3}{5k^2}\left(\dfrac{\mg}{m_3} \right) \right]^2
\eea
and 
\bea
&\Omega_{3/2}h^2 \approx 5.9\times10^{5}\dfrac{k^4}{\lambda^{10}}\left(\dfrac{\alpha_3}{0.1}\right)^{-2}\left(\dfrac{m_3}{1\TeV}\right)^2\left(\dfrac{\mg}{1\GeV} \right)^{-2}
\nonumber\\
&\times\left(\dfrac{m}{10^{14}\GeV}\right)^5\left(\dfrac{T_R}{10^8\GeV}\right)\left[1+\dfrac{12\pi\alpha_3}{5k^2}\left(\dfrac{\mg}{m_3} \right) \right]^2
\eea
The reheating temperature $T_R$ required for the correct density of gravitino dark matter is then 
\bea
&T_R\approx 17 \GeV\times\dfrac{\lambda^{10}}{k^4}\left(\dfrac{\alpha_3}{0.1}\right)^{2}
\left(\dfrac{m_3}{1\TeV}\right)^{-2}
\nonumber\\
&\times\left(\dfrac{\mg}{1\GeV} \right)^{2}\left(\dfrac{m}{10^{14}\GeV}\right)^{-5}\left[1+\dfrac{12\pi\alpha_3}{5k^2}\left(\dfrac{\mg}{m_3} \right) \right]^{-2}\,.
\label{reheating12}
\eea
Eqs.(\ref{reheating21}) and (\ref{reheating12}) are the main results of this article. We see that the model can accommodate a very wide range of $T_R$ and still be consistent with gravitino dark matter from the GMSB sector.
The most striking feature of these results is their extreme sensitivity to the parameters of the model, in particular the \OR sector coupling $\lambda$. This means that the gravitino density from decay of the SUSY breaking scalar in the GMSB sector can account for dark matter for essentially any value of the reheating temperature above the BBN bound, $T_{R} \gsim 1 \MeV$. 

  Comparing our results with the previous results of \cite{ET}, which assumed that oscillations about the minimum began at $H_{osc} \approx m_{S}$ with $S_{osc} \approx \sqrt{2} c$ (for real $S$), we find:  
\bea
&\dfrac{n_{3/2}}{s}\approx\left.\dfrac{n_{3/2}}{s}\right|_{ET}\times \left(\dfrac{a}{c} \right)^3\dfrac{1}{m_S^6}\,,\qquad \left|\dfrac{a}{cm_S^2}\right|<1\,,\\
&\dfrac{n_{3/2}}{s}\approx\left.\dfrac{n_{3/2}}{s}\right|_{ET}\times \left(\dfrac{a}{c} \right)^2\dfrac{1}{m_S^4}\,,\qquad \left|\dfrac{a}{cm_S^2}\right|>1\,,
\eea
where $\left.n_{3/2}/s\right|_{ET}$ is the value given in \cite{ET},
\bea &\left.\dfrac{n_{3/2}}{s}\right|_{ET} \approx 
2 \times 10^{-10} k^{4} \left(\dfrac{\alpha_{3}}{0.1}\right)^{2} \left(\dfrac{m_{3}}{1 \TeV}\right)^{-2} 
\nonumber\\ 
&\times \left(\dfrac{m_{3/2}}{1 \GeV}\right) \left(\dfrac{T_{R}}{10^{8} \GeV}\right)
 \left(\dfrac{\Lambda}{10^{14} \GeV}\right) ~.\eea
If $\Hub_1 > \Hub_2$, $|a/c m_S^2|>1$ and so there is a strong enhancement of the gravitino abundance relative to the amplitude obtained in \cite{ET}. Similarly, if $H_2 > H_1$ then there is a strong suppression of the gravitino abundance relative to \cite{ET}. 

     Note that while in the existing literature $T_R$ is necessarily written in terms of an arbitrary cutoff \cite{ET}, in our analysis there is a fully defined perturbative hidden sector, eliminating the cut-off in favour of the masses and couplings of the model. Therefore what could have been a limitation, namely focusing on a specific \OR-type sector, is in fact an advantage. Since there is no cut-off, there is no need to constrain the parameters of the model in order to make it consistent with the validity of the effective theory.

\section{Conclusions and outlook}

			In this paper we have considered the dynamics of the SUSY breaking scalar $S$ in a GMSB scenario with metastable vacua and the production of gravitino dark matter through its decay. Our results for the cosmological evolution of the scalar field and the resulting gravitino density are significantly different from previous investigations.
						
			We have shown that since the Universe is expanding, the minimum of the potential $S_0$ is time-dependent. The $S$ field tracks the minimum, which generates a displacement $\delta S$. Once the minimum becomes $H$ independent, $S$ begins coherent oscillations about the minimum with initial amplitude determined by the displacement.  This produces a very different $S$ oscillation amplitude and so gravitino density as compared with previous analyses \cite{ET}. By considering a generic potential $V(S)$, we have shown that there are two possible values of $\delta S$, depending on whether the quadratic or the linear term first becomes $H$-independent.
The resulting gravitino density can be highly suppressed or enhanced as compared with the previous estimate of \cite{ET}, depending on the parameters of the model. A striking feature of the gravitino density is its extreme sensitivity to the parameters of the model, with the reheating temperature having a $\lambda^{10}$ or $\lambda^{14}$ dependence on the superpotential coupling of the \OR sector.  As a result, it is easy to account for gravitino dark matter with an arbitrarily low reheating temperature. (There are also constraints on $m_{3/2}$ and $\lambda$ from gravitino free-streaming, but as shown in Appendix C these are easily satisfied.) 

 This could be significant for the cosmology of GMSB models. It is possible that the NLSP could be discovered at the LHC and its properties established. In particular, the thermal freeze-out density and decay rate of the NLSP may turn out to be inconsistent with BBN. Should the LHC discover a cosmologically problematical NLSP, a low reheating temperature would be the simplest way to solve the problem, by suppressing the NLSP density via entropy  
release prior to $T_{R}$. This requires that $T_{R}$ is sufficiently small compared with the 
freeze-out temperature of the NLSP, $T_{R} \ll M_{NLSP}/20 \lsim 50 \GeV$. In this case the GMSB sector would be a prime candidate for the origin of the gravitino dark matter density in the presence of a low $T_{R}$. (However, the reheating temperature required to produce the correct density of gravitino dark matter is very sensitive to the parameters of the model, so obtaining a low enough reheating temperature to dilute the NLSP density while not affecting BBN may require a degree of coincidence.) To complete the model a source of baryogenesis consistent with a low $T_{R}$ would be required. A natural possibility would be Affleck-Dine baryogenesis. Indeed, for $m_{3/2} \sim 1 \GeV$, Affleck-Dine baryogenesis combined with Q-ball decay in GMSB could simultaneously account for both gravitino dark matter and the baryon asymmetry \cite{shoe}, but for smaller $m_{3/2}$ it would provide only the 
baryon asymmetry.

    As an additional remark, we note that the \OR GMSB sector is a well-defined perturbative model. Therefore the decay of the SUSY-breaking scalar field is addressed without any cutoff, nor with any stringent constraints on the mass scales nor on the coupling constants.

   \section*{Acknowledgements} 
Support of the European Union, through the Marie Curie Research and Training Network "UniverseNet" (MRTN-CT-2006-035863), is appreciated. The work of AF was also supported by the Academy of Finland, grant 114419. The authors would like to thank Kari Enqvist for observations and comments, and Kazunori Kohri for his useful suggestions. AF would also like to thank the staff at the University of Lancaster (UK), for the kind hospitality in various occasions during the course of this work.

\appendix

\section{Radiative corrections at one loop and effective potentials}

In this section we calculate the radiative corrections to the scalar potential $V(S)$ and to the \K potential of $S$. The supergravity contribution is obtained accordingly.

\subsection{Contribution of global SUSY}
 
The potential $V_{SUSY}$ includes the classical contribution $V_0$ and the effective potential $V_{eff}$ that is generated by loops of $X$, $Y$ and of the messengers $f$ and $\bar{f}$,
\bea
&V_{SUSY}(S)=V_0(S)+V_{eff}(S)\equiv V^h_0(S)+V^m_0(S)\nonumber\\
&+V^h_{eff}(S)+V^m_{eff}(S)\,.
\label{sp}
\eea
The classical scalar potential of the hidden sector, $V^h_0$ has the following form
\bea
&V^h_0(S)=\mu^4+\lambda^2|X|^4-\lambda\mu^2(X^2+X^{\dagger 2})+m^2(|X|^2+|Y|^2)\nonumber\\
&+2\lambda m(SXY^{\dagger}+S^{\dagger} X^{\dagger} Y)+4\lambda^2|X|^2|S|^2\,,
\label{vclassic}
\eea
which is given by the $F$-terms derived from Eq.(\ref{super}).

In general, the scalar and fermion loops correspond to the Coleman-Weinberg potentials \cite{CW}
\be
V_{eff}^\phi=\dfrac{1}{64\pi^2}\mathrm{Tr}\left[M_\phi^2\ln\dfrac{M_\phi}{R}\right]\,,
\label{cws}
\ee
for the scalars, and
\be
V_{eff}^\psi=-\dfrac{1}{64\pi^2}\mathrm{Tr}\left[(|M_\psi|^2)^2\ln\dfrac{|M_\psi|^2}{R^2}\right]\,,
\label{cwf}
\ee
for the fermions. The minus sign comes as usual from the Fermi-Dirac statistics. The mass matrices $M_\psi$ and $M_\phi$ are given by equations (\ref{fmm}) and (\ref{smm}) respectively, and $R$ is an ultraviolet cutoff.

The formulas (\ref{cws}) and (\ref{cwf}) give the following Coleman-Weinberg potential for the hidden sector,
\bea
&&V^h_{eff}(S)=V^h_\phi(S)+2V^h_\psi(S)\approx
\nonumber\\
&&\approx\dfrac{\lambda^4\mu^4}{3\pi^2m^2}|S|^2-\dfrac{3\lambda^6\mu^4}{10\pi^2m^4}|S|^4+\ldots
\label{susyeff}
\eea
where we have dropped a constant and kept the dominant terms in $\lambda\mu^2/m^2$ and in $|S|^2$. The above equation agrees with \cite{MN}. The positive mass term, which for small values of $S$ dominates over the quartic term, confirms what has been claimed in Section \ref{cosmo}.
By adding now Eq.(\ref{susyeff}) to (\ref{vclassic}), the scalar potential of the hidden sector can be recast as follows,
\bea
&V^h(S)=V^h_0(S)+V^h_{eff}(S)\approx \nonumber\\
&\approx\mu^4+\lambda^2|X|^4-\lambda\mu^2(X^2+X^{\dagger 2})+m^2(|X|^2+|Y|^2)+\nonumber\\
&+2\lambda m(SXY^{\dagger}+S^{\dagger} X^{\dagger}Y)+4\lambda^2|X|^2|S|^2+\nonumber\\ &+\dfrac{\lambda^4\mu^4}{3\pi^2m^2}|S|^2-\dfrac{3\lambda^6\mu^4}{10\pi^2m^4}|S|^4\,.
\label{totalbasic}
\eea
Considering now the messenger sector, one can act in complete analogy. The $F$-terms of the messengers give the following tree-level potential:
\be
V_0^m=k(-\mu^2+\lambda X^{\dagger 2})f\bar{f}+k(-\mu^2+\lambda X^2)f^{\dagger}\bar{f}^{\dagger}+k^2|f|^2|\bar{f}|^2\,.
\label{classmess}
\ee
The mass matrices of the messenger scalars and fermions are given by equations (\ref{msmm}) and (\ref{fmmm}) in Section \ref{cosmo}. By summing the two contributions as before, and keeping only the dominant terms in $k\mu^2/M^2$, in $\lambda\mu^2/m^2$ and in $S$, we obtain the Coleman-Weinberg potential of the messengers,
\bea
&V_{eff}^m(S)\approx 
\dfrac{5k^3\mu^2}{16\pi^2}\left[\dfrac{\mu^2}{M}-\dfrac{\lambda}{M}(X^2+X^{\dagger 2})\right](S+S^{\dagger})-
\nonumber\\
&-\dfrac{5k^4\mu^4}{32\pi^2M^2}(S^2+S^{\dagger 2})+\dfrac{5k^6\mu^4}{24\pi^2M^6}\bigg [\dfrac{\mu^4}{2}+\dfrac{\lambda^2}{2}(X^4+X^{\dagger 4}+
\nonumber\\
&+4|X|^4)+\lambda \mu^2(X^2+X^{\dagger 2})\bigg]|S|^2\,,
\label{cwm}
\eea
where we have dropped an unimportant constant. The factor of 5 comes from representations of SU(5). This expression agrees with the literature \cite{ET,MN} for $X=Y=0$. Notice that the messenger loops generate a mass term for the field $S$ as well.

By adding (\ref{classmess}) and (\ref{cwm}), the scalar potential of the hidden and messenger sector along the direction $X=Y=f=\bar{f}=0$ in the field space, which is our case of interest, can be written as
\bea
&V_{SUSY}(S)\approx \dfrac{5k^3\mu^4}{16\pi^2M}(S+S^{\dagger})-\dfrac{5k^4\mu^4}{32\pi^2M^2}(S^2+S^{\dagger 2})+\nonumber\\
&+\left(\dfrac{\lambda^4\mu^4}{3\pi^2m^2}+\dfrac{5k^6\mu^8}{48\pi^2M^6}\right)|S|^2+\dfrac{5k^7\mu^8}{32\pi^2M^7}(S+S^{\dagger})|S|^2-\nonumber\\
&-\left(\dfrac{3\lambda^6\mu^4}{10\pi^2m^4}+\dfrac{15k^4\mu^8}{64\pi^2M^8}\right)|S|^4\,.
\label{potglob}
\eea
This will be added to the supergravity corrections of the next section, in order to obtain the full potential of the model.

\subsection{\K and supergravity corrections to the potential}

Let us first focus on the \K potential. In the case at hand, Eq.(\ref{masscorr}) is rewritten as $\mu_{ik}=(1-\delta_{ik})(M+kS)$ for the messengers. Therefore Eq.(\ref{Kcorr}) can be rewritten as follows,
\be
K^{(1)}_m=-\dfrac{5}{32\pi^2}\mathrm{Tr}\left[|M^m_\psi|^2\left(\ln\dfrac{|M^m_\psi|^2}{\Lambda^2}-1\right)\right]\,,
\ee
where, as before, the factor of 5 corresponds to SU(5) representations. At the lowest order in $k\mu^2/M^2$ and for small values of $S$, we obtain
\bea
&K^{(1)}_m(S)\approx-\dfrac{5M^2}{16\pi^2}\left[\dfrac{k}{M}(S+S^{\dagger})+2\left(\dfrac{k}{M}\right)^2|S|^2+\right.\nonumber\\
&\left.+\dfrac{1}{2}\left(\dfrac{k}{M}\right)^3|S|^2(S+S^{\dagger})-\dfrac{1}{6}\left(\dfrac{k}{M}\right)^4|S|^2(S^2+S^{\dagger 2})\right]\,,\nonumber\\
\label{cwkahler}
\eea
as it was found also in \cite{ET}. On the other hand, loops of the superfields $X$ and $Y$ in the hidden sector generate the following \K potential, expanded in $S$ and in $\lambda\mu^2/m^2$,
\be
K^{(1)}_h(S)\approx-\dfrac{\lambda^2}{8\pi^2}|S|^2-\dfrac{\lambda^4}{12\pi^2m^2}|S|^4+\cdots
\label{xykahler}
\ee
The above agrees with the statement reported in \cite{MN}, namely these corrections correspond to the \K potential
\be
K \approx |S|^2-\dfrac{\lambda^2}{8\pi^2}|S|^2-\dfrac{\lambda^4}{12\pi^2m^2}|S|^4\approx|S|^2-\dfrac{|S|^4}{4\Lambda^2}+\ldots
\label{kc}
\ee
with
\be
\Lambda^2=\dfrac{3\pi^2m^2}{\lambda^4}\,,
\ee
since $\lambda^2/8\pi^2\ll 1$ even for a strong coupling $\lambda\approx\mathcal{O}(1)$.

To embed the model in supergravity, we can use the general formula \cite{Nilles}
\bea
&V_{Sugra}  =  \exp(K/\mpl)  \left[ \sum_{\alpha,\beta}
\left( \dfrac{\partial^2 K}{\partial \bar{\phi}_{\alpha} \partial \phi_{\beta} }
	\right)^{-1}\times\right.\nonumber\\
	 &\left.\times
\left( \dfrac{\partial W}{\partial \phi_{\alpha} } + \dfrac{W}{\mpl}
	 \dfrac{\partial K}{\partial \phi_{\alpha} } \right)
	 \left( \dfrac{\partial \overline{W} }{\partial \bar{\phi}_{\beta} }
	+ \dfrac{\overline{W}}{\mpl}
	\dfrac{\partial K}{\partial \bar{\phi}_{\beta} } \right)
- 3 \dfrac{|W|^2}{\mpl}  \right]\,,\nonumber\\
\label{sugra}
\eea
with the \K derivative of the superpotential $W$
\be
DW\equiv\left( \dfrac{\partial W}{\partial \phi_{\alpha} } + \dfrac{W}{\mpl}
	 \dfrac{\partial K}{\partial \phi_{\alpha} } \right)\,.
\ee
At this point, it might be instructive to show the effect of the quartic term in $S$ in the \K potential at the tree level, before including the radiative corrections from the loops of the messengers and of the hidden sector.
Eq.(\ref{sugra}) for the SUSY breaking field is rewritten in our case as
\bea
&V_{Sugra}(S)  =  \mu^4 \exp(K/\mpl)  \left[\left(1-\dfrac{|S|^2}{\Lambda^2}\right)^{-1}\times\right.\nonumber\\
&\left.\times\left(1+2\dfrac{|S|^2}{\mpl}+\dfrac{|S|^4}{M_{Pl}^4} \right)- 3 \dfrac{|S|^2}{\mpl}  \right]\,,
\eea
with the \K potential (\ref{kc}). At the lowest order in $|S|^2$, this is expanded as
\bea
&V_{Sugra}(S) \approx \mu^4\left( 1+\dfrac{|S|^2}{\Lambda^2}+\dfrac{7}{4}\dfrac{|S|^4}{\Lambda^2\mpl}\right)\nonumber\\
&
=\mu^4\left( 1+\dfrac{\lambda^4}{3\pi^2}\dfrac{|S|^2}{m^2}+\dfrac{7\lambda^4}{12\pi^2}\dfrac{|S|^4}{m^2\mpl}\right)\,.
\label{sugracorr}
\eea
The term proportional to $|S|^2$ has a positive coefficient, consistently with the literature \cite{ET}.

If the radiative corrections (\ref{cwkahler}) and (\ref{xykahler}) are now taken into account, the dominant terms are
\bea
&K^{(1)}(S)=K^{(1)}_m(S)+K^{(1)}_h(S)\approx  |S|^2-\dfrac{5k}{16\pi^2}M(S+S^{\dagger})-\nonumber\\
&-\dfrac{5k^3}{32\pi^2}\dfrac{|S|^2}{M}(S+S^{\dagger})-\dfrac{\lambda^4}{12\pi^2}\dfrac{|S|^4}{m^2}+\ldots
\eea
thus the following dominant terms are generated,
\bea
&\hspace{-1cm}V_{Sugra}(S) \approx  \mu^4\dfrac{|S|^2}{\mpl}\left\{\dfrac{25k^2}{64\pi^4}\left[-k^2+\left(\dfrac{M}{M_{Pl}}\right)^2\right]\right.\nonumber\\
&\hspace{-0.5cm}\left.+\dfrac{\lambda^4}{3\pi^2}\left(\dfrac{M_{Pl}}{m}\right)^2\right\}+\dfrac{5k\mu^4}{16\pi^2}(S+S^{\dagger})\left( \dfrac{k^2}{M}-2\dfrac{M}{\mpl}\right)+\ldots\nonumber\\
\eea
where we write only the terms which are linear and quadratic in $S$. This agrees with (\ref{sugracorr}) in the limit $k\rightarrow 0$. We can further assume that 
\be
\dfrac{25k^2}{64\pi^4}\left|-k^2+\left(\dfrac{M}{M_{Pl}}\right)^2\right|\ll\dfrac{\lambda^4}{3\pi^2}\left(\dfrac{M_{Pl}}{m}\right)^2,
\ee
therefore the dominant supergravity corrections can be rewritten as
\be
V_{Sugra}(S) \approx \mu^4\left[ \dfrac{\lambda^4}{3\pi^2m^2}|S|^2+\dfrac{5k}{16\pi^2}(S+S^{\dagger})\left( \dfrac{k^2}{M}-2\dfrac{M}{\mpl}\right)
\right].
\label{vsugra}
\ee
Clearly, the sign of the term that is linear in $S$ is now determined by the coupling constant $k$. In particular, a form of the potential such as
\be
V_{Sugra}(S) \approx \mu^4\left[ +c^2|S|^2-d^2(S+S^{\dagger})+\ldots\right]\,,
\ee
would be recovered if the following condition is satisfied:
\be
k<\sqrt{2}\dfrac{M}{M_{Pl}}\Rightarrow\dfrac{\mg}{m_3}>\dfrac{5.77}{\alpha_3}\,.\label{inc}
\ee
The above equation directly relates the gravitino mass $\mg$ to the gluino mass $m_3$ and to the strong coupling constant $\alpha_3$. It is calculated by using Eq.(\ref{massM}) in Section \ref{cosmo}. However, (\ref{inc}) is not consistent with gauge mediation, where the gravitino is the LSP. In particular, the framework that is used in this paper allows gravitino masses $\mg\lsim 10 \GeV$, thus we conclude that also the coefficient of the linear term in (\ref{vsugra}) is identically positive.

The supergravity corrections to the scalar potential of the model can now be written by adding (\ref{vsugra}) to the potential for $S$ which is generated by the inflaton, whose energy density dominates the Universe before the field starts oscillating \cite{ET},
\be
V(S) \approx e^{K} (3 H^2 M_P^2)\approx 3 H^2 \left[ |S|^2 - \dfrac{5 k}{16 \pi^2} M(S+S^\dagger) + \cdots\right]\,,
\label{V}
\ee
(where K has been normalised in units of Planck mass). This provides, at the lowest order in the SUSY breaking field,
\bea
&V_{Sugra}(S) \approx 3 H^2 \left[ |S|^2 - \dfrac{5 k}{16 \pi^2} M(S+S^\dagger)\right]+\nonumber\\
&+\mu^4\left[\dfrac{\lambda^4}{3\pi^2m^2}|S|^2+\dfrac{5k}{16\pi^2}(S+S^{\dagger})\left( \dfrac{k^2}{M}-2\dfrac{M}{\mpl}\right)\right]\,,\nonumber\\
\label{sugrafinal}
\eea
namely with the supergravity corrections which, together with Eq.(\ref{potglob}), are used in Section \ref{cosmo} to calculate the gravitino abundance and the reheating temperature.

\section{Consistency of the Coleman Weinberg potentials and constraints on the gravitino mass range}

In this appendix we first note that the coupling constants of our model $\lambda$ and $k$ and the mass scale $m$ are constrained by requiring consistency of the series expansion of the Coleman-Weinberg potential.
The scalar potential Eq.(\ref{total}) is obtained by expanding in terms of $\lambda\mu^2/m^2$, so the condition
\be
\dfrac{\lambda\mu^2}{m^2}\ll 1\,,
\label{mmu}
\ee
must be satisfied. Therefore 
\be m > \lambda^{1/2} \mu\,.
\label{ratio}
\ee
Next we derive further constraints on the mass scale $m$ of the hidden sector, and show that the gravitino mass range which is consistent with the model can be accordingly constrained. First, using \eq{mu2}, the lower bound (\ref{ratio}) can be recast as
\be
m > \lambda^{1/2} \mu=2\times10^{9} \lambda^{1/2} \GeV\sqrt{\dfrac{\mg}{1 \GeV}}\,.
\label{lowerm}
\ee
In Section \ref{cosmo}, we assume that the gravitino is the main decay product of the field $S$ when this rapidly oscillates around the minimum. Accordingly, the upper limit Eq.(\ref{mdecay}) holds:
\be
m\lsim 1.5 \times 10^{14} \; \lambda^2 \GeV \times\sqrt{\dfrac{\alpha_3}{0.1}}\left(\dfrac{m_3}{1\TeV} \right)^{-1}\left(\dfrac{\mg}{1\GeV} \right).
\label{upperm}
\ee
Clearly, (\ref{lowerm}) and (\ref{upperm}) must be simultaneously satisfied. This means that the formulas (\ref{reheating21}) and (\ref{reheating12}) are valid only for certain gravitino mass ranges, depending on the values of $\lambda$. In particular,
\bea
&\qquad \lambda\approx 10^{-2}\Rightarrow 1\MeV\lsim\mg\lsim 10\GeV\,,\label{mgconstr1}\\
&\qquad \lambda\approx 10^{-1}\Rightarrow 1\keV\lsim\mg\lsim 10\GeV\,,
\label{mgconstr2}
\eea
where we have assumed only that $m_3\approx 1\TeV$. 
Remarkably, since these results follow from the consistency of the Coleman-Weinberg potentials and from phenomenology, they do not rely on any approximations.

Let us finally show why the coupling $k$ can be consistently of order unity.
In analogy with Eq.(\ref{ratio}), the effective potential (\ref{cwm}) for the messenger sector is valid if
\be
\dfrac{k\mu^2}{M^2}\ll 1\,.
\label{expk}
\ee
By recalling Eq.(\ref{massM}), i.e.
\be
M \approx 3.3 \times 10^{13} \GeV \times k \left(\dfrac{\alpha_3}{0.1} \right)\left(\dfrac{1\TeV}{m_3} \right)\left(\dfrac{\mg}{1\GeV} \right)\,,
\ee
we can substitute the above equation and (\ref{mu2}) in (\ref{expk}) to finally obtain
\be
\dfrac{10^{-8}}{2k}
\left( \frac{0.1}{\alpha_{3}} \right)^{2} 
\left( \frac{m_{3}}{1 \TeV} \right)^{2}
\left(\dfrac{\mg}{1\GeV} \right)^{-1}\ll 1\,.
\ee 
This condition is satisfied for $k \gsim 0.01$ in the entire gravitino mass range in equations (\ref{mgconstr1}) and (\ref{mgconstr2}).

\section{Constraints from gravitino free-streaming}

	One should also take into account the free streaming length $\lambda_{FS}$, namely the distance that the gravitinos produced through the decay can travel until they become non-relativistic. If we want them to be Cold Dark Matter (CDM), this quantity must be smaller than $\mathcal{O}(100 \,\rm kpc)$.

 The free-streaming length for gravitinos from $S$ decay was derived in \cite{ET}, 
\be \lambda_{FS} \approx 1 {\rm\, kpc} \sqrt{\dfrac{g_*}{100}}   \left(\dfrac{m_{\g}}{1{\rm GeV}}\right)^{-\frac{3}{2}}
\left(\frac{\Lambda}{10^{14} \GeV}\right)^{3/2}    ~,\ee
where $g_*$ is the number of relativistic degrees of freedom at the time of decay.
Using $\Lambda = \sqrt{3} \pi m/\lambda^2$, this becomes
\be
\lambda_{FS}\approx 1 {\rm\, kpc}\, \left(\dfrac{\sqrt{3}\pi}{\lambda^2} \right)^{\frac{3}{2}}  \sqrt{\dfrac{g_*}{100}}   \left(\dfrac{m_{\g}}{1{\rm GeV}}\right)^{-\frac{3}{2}}\left(\dfrac{m}{10^{14} {\rm GeV}}\right)^{\frac{3}{2}}\,.
\ee
Since the free streaming length must be smaller than 100 kpc, one obtains a lower limit on the gravitino mass,
\be
m_{\g}\gsim 0.25\times\dfrac{1}{\lambda^2}\left(\dfrac{g_*}{100}\right)^{\frac{1}{3}}\left(\dfrac{m}{10^{14} {\rm GeV}}\right)\GeV\,.
\label{fslmass}
\ee
This depends on $m$ and $\lambda$. Using the lower bound on $m$, Eq.(\ref{lowerm}), we obtain a lower bound on the gravitino mass from the requirement that gravitinos behave like cold dark matter
\bee{bound} m_{3/2} \gae 2.6 \times 10^{-2} \eV \times \frac{1}{\lambda^{3}} \left(\dfrac{g_*}{100}\right)^{\frac{2}{3}}\; 
~.\eee 
From this we see that free-streaming imposes only a weak constraint on the gravitino mass, unless $\lambda\ll 1$.



\begin{thebibliography}{99}
  

\bibitem{GMSB}
 M.~Dine, W.~Fischler and M.~Srednicki,
 Nucl.\ Phys.\ B {\bf 189}, 575 (1981);
 S.~Dimopoulos and S.~Raby,
 Nucl.\ Phys.\ B {\bf 192}, 353 (1981);
 M.~Dine and W.~Fischler,
 Phys.\ Lett.\ B {\bf 110}, 227 (1982);
 M.~Dine and W.~Fischler,
 Nucl.\ Phys.\ B {\bf 204}, 346 (1982);
 C.~R.~Nappi and B.~A.~Ovrut,
 Phys.\ Lett.\ B {\bf 113}, 175 (1982);


\bibitem{GR}
 For a review, see G.~F.~Giudice and R.~Rattazzi,
  Phys.\ Rept.\  {\bf 322} (1999) 419
  [arXiv:hep-ph/9801271].
 

\bibitem{gravitino}  
M.~Lemoine, G.~Moultaka and K.~Jedamzik,
  Phys.\ Lett.\  B {\bf 645} (2007) 222
  [arXiv:hep-ph/0504021];
  K.~Jedamzik, M.~Lemoine and G.~Moultaka,
  Phys.\ Rev.\  D {\bf 73} (2006) 043514
  [arXiv:hep-ph/0506129];
  K.~Jedamzik,
  Phys.\ Rev.\  D {\bf 74} (2006) 103509
  [arXiv:hep-ph/0604251];
  S.~Bailly, K.~Y.~Choi, K.~Jedamzik and L.~Roszkowski,
  JHEP {\bf 0905} (2009) 103
  [arXiv:0903.3974 [hep-ph]].
 

\bibitem{ellis}
J.~R.~Ellis, J.~E.~Kim and D.~V.~Nanopoulos,
  Phys.\ Lett.\  B {\bf 145}, 181 (1984).





\bibitem{BBB}
  M.~Bolz, A.~Brandenburg and W.~Buchmuller, 
  Nucl.\ Phys.\ {\bf B 606} (2001) 518
  [Erratum-ibid.\ {\bf  B 790} (2008) 336]
  [hep-ph/0012052];
  
  
   
\bibitem{PS}
  
  J.~Pradler and F.~D.~Steffen,
  Phys.\ Rev.\ {\bf D 75} (2007) 023509 [hep-ph/0608344].
 


\bibitem{Andrea}
  A.~Ferrantelli,
  JHEP {\bf 0901} (2009) 070
  [arXiv:0712.2171 [hep-ph]].



\bibitem{Narendra}
  R.~Rangarajan and N.~Sahu,
  Phys.\ Rev.\  D {\bf 79} (2009) 103534
  [arXiv:0811.1866 [hep-ph]].
   


\bibitem{infldecay}

  M.~Kawasaki, F.~Takahashi and T.~T.~Yanagida,
  Phys.\ Lett.\  B {\bf 638} (2006) 8
  [arXiv:hep-ph/0603265];
  T.~Asaka, S.~Nakamura and M.~Yamaguchi,
  Phys.\ Rev.\  D {\bf 74} (2006) 023520
  [arXiv:hep-ph/0604132].


\bibitem{moduli}
  M.~Endo, K.~Hamaguchi and F.~Takahashi,
  Phys.\ Rev.\ Lett.\  {\bf 96} (2006) 211301
  [arXiv:hep-ph/0602061];
  S.~Nakamura and M.~Yamaguchi,
  Phys.\ Lett.\  B {\bf 638} (2006) 389
  [arXiv:hep-ph/0602081].
  
  
\bibitem{Dine}

  M.~Dine, R.~Kitano, A.~Morisse and Y.~Shirman,
  Phys.\ Rev.\  D {\bf 73} (2006) 123518
  [arXiv:hep-ph/0604140].
  


\bibitem{Coughlan}
  G.~D.~Coughlan, R.~Holman, P.~Ramond and G.~G.~Ross,
  Phys.\ Lett.\  B {\bf 140} (1984) 44.

\bibitem{Banks}
 T.~Banks, D.~B.~Kaplan and A.~E.~Nelson,
 Phys.\ Rev.\  D {\bf 49}, 779 (1994)
 [arXiv:hep-ph/9308292].
   
    
   
\bibitem{IK}
  M.~Ibe and R.~Kitano,
  Phys.\ Rev.\  D {\bf 75} (2007) 055003
  [arXiv:hep-ph/0611111].

   
\bibitem{ET}
  M.~Endo and F.~Takahashi,
  arXiv:0710.1561 [hep-ph].
   

      
   \bibitem{HKT}
  K.~Hamaguchi, R.~Kitano and F.~Takahashi,
  JHEP {\bf 0909} (2009) 127
  [arXiv:0908.0115 [hep-ph]].



  
   
   
      
\bibitem{Kaz}
  M.~Kawasaki, K.~Kohri, T.~Moroi and A.~Yotsuyanagi,
 \textit{ Phys.\ Rev.\ }{\bf D 78} (2008) 065011
  [arXiv:0804.3745];
  L.~Covi, J.~Hasenkamp, S.~Pokorski and J.~Roberts,
  JHEP {\bf 0911} (2009) 003
  [arXiv:0908.3399 [hep-ph]].

   
\bibitem{Felder}
 G.~N.~Felder, L.~Kofman and A.~D.~Linde,
 Phys.\ Rev.\  D {\bf 60}, 103505 (1999)
 [arXiv:hep-ph/9903350].
 
\bibitem{GTA}
 G.~F.~Giudice, I.~Tkachev and A.~Riotto,
 JHEP {\bf 9908}, 009 (1999)
 [arXiv:hep-ph/9907510].


\bibitem{Maroto}
 A.~L.~Maroto and A.~Mazumdar,
 Phys.\ Rev.\ Lett.\  {\bf 84}, 1655 (2000)
 [arXiv:hep-ph/9904206].


\bibitem{Kawasaki}
  M.~Kawasaki, F.~Takahashi and T.~T.~Yanagida,
  Phys.\ Rev.\  D {\bf 74}, 043519 (2006)
  [arXiv:hep-ph/0605297].
  
   
   

\bibitem{Jedamzik}
K.~Jedamzik and M.~Pospelov,
  New J.\ Phys.\  {\bf 11} (2009) 105028
  [arXiv:0906.2087 [hep-ph]].


   
   
   
   
   
   
  
\bibitem{T}
  M.~S.~Turner,
  Phys.\ Rev.\  D {\bf 28} (1983) 1243;
  J.~E.~Kim,
  Phys.\ Rept.\  {\bf 150} (1987) 1;
  K.~A.~Olive,
  Phys.\ Rept.\  {\bf 190} (1990) 307;
 N.~Bartolo, E.~Komatsu, S.~Matarrese and A.~Riotto,
  Phys.\ Rept.\  {\bf 402} (2004) 103
  [arXiv:astro-ph/0406398].
   
     
  \bibitem{BBN}
G.~D.~Coughlan, W.~Fischler, E.~W.~Kolb, S.~Raby and G.~G.~Ross,
Phys.\ Lett.\  B {\bf 131}, 59 (1983).




  
   \bibitem{gravover}
  M.~Dine, W.~Fischler and D.~Nemeschansky,
  Phys.\ Lett.\  B {\bf 136} (1984) 169.



  
  
    \bibitem{Lowreh}
 
 See for instance G.~F.~Giudice, E.~W.~Kolb and A.~Riotto,
  Phys.\ Rev.\ D  {\bf64} (2001) 023508
  [hep-ph/0005123];
  G.~F.~Giudice, E.~W.~Kolb, A.~Riotto, D.~V.~Semikoz and I.~I.~Tkachev,
  Phys.\ Rev.\  D {\bf 64}, 043512 (2001)
  [arXiv:hep-ph/0012317];
  N.~Fornengo, A.~Riotto and S.~Scopel,
  Phys.\ Rev.\  D {\bf 67}, 023514 (2003)
  [arXiv:hep-ph/0208072];
M.~Kawasaki, K.~Kohri and T.~Moroi,
  Phys.\ Rev.\ D {\bf 71} (2005) 083502
  [astro-ph/0408426].
  
  
  
  

    
\bibitem{MN} H.~Murayama and Y.~Nomura,
  Phys.\ Rev.\  D {\bf 75} (2007) 095011
  [arXiv:hep-ph/0701231].
  
  
  
    




\bibitem{therm}

S.~A.~Abel, C.~S.~Chu, J.~Jaeckel and V.~V.~Khoze,
JHEP {\bf 0701}, 089 (2007)
[arXiv:hep-th/0610334];
N.~J.~Craig, P.~J.~Fox and J.~G.~Wacker,
Phys.\ Rev.\ D {\bf 75}, 085006 (2007)
[arXiv:hep-th/0611006];
W.~Fischler, V.~Kaplunovsky, C.~Krishnan, L.~Mannelli and M.~A.~C.~Torres,
JHEP {\bf 0703}, 107 (2007)
[arXiv:hep-th/0611018].






  
\bibitem{CW}
  S.~R.~Coleman and E.~Weinberg,
  Phys.\ Rev.\  D {\bf 7} (1973) 1888.

  
\bibitem{Grisaru}
  M.~T.~Grisaru, M.~Rocek and R.~von Unge,
  Phys.\ Lett.\  B {\bf 383} (1996) 415
  [arXiv:hep-th/9605149].  
  
  





  

  

  
  
  


  
\bibitem{Nilles}
  H.~P.~Nilles,
  Phys.\ Rept.\  {\bf 110} (1984) 1.

  

\bibitem{Lyth}
  E.~J.~Copeland, A.~R.~Liddle, D.~H.~Lyth, E.~D.~Stewart and D.~Wands,
  Phys.\ Rev.\  D {\bf 49} (1994) 6410
  [arXiv:astro-ph/9401011].
  


  \bibitem{shoe}
  I.~M.~Shoemaker and A.~Kusenko,
  Phys.\ Rev.\  D {\bf 80} (2009) 075021
  [arXiv:0909.3334 [hep-ph]].
 
  









  
\end{thebibliography}
\end{document}